# Reframing the Test Pyramid
# for Digitally Transformed Organizations

Nicole Radziwill & Graham Freeman

**Abstract**: The test pyramid is a conceptual model that describes how quality checks can be organized to ensure coverage of all components of a system, at all scales. Originally conceived to help aerospace engineers plan tests to determine how material changes impact system integrity, the concept was gradually introduced into software engineering. Today, the test pyramid is typically used to illustrate that the majority of tests should be performed at the lowest (unit test) level, with fewer integration tests, and even fewer acceptance tests (which are the most expensive to produce, and the slowest to execute). Although the value of acceptance tests and integration tests increasingly depends on the integrity of the underlying data, models, and pipelines, software development and data management organizations have traditionally been siloed and quality assurance practice is not as mature in data operations as it is for software. Companies that close this gap by developing cross-organizational systems will create new competitive advantage and differentiation. By taking a more holistic view of testing that crosses these boundaries, practitioners can help their organizations close the gap.

**Keywords:** Software testing, test pyramid, automation, test automation, data testing, Industry 4.0

## Introduction

In 2016, research firm Forrester identified the "insights-driven business" as a new type of company that is emerging during the age of digital innovation. (Fenty, 2016) These firms develop strategic insights using data science and software supported by closed-loop systems, which creates significant differentiation and competitive advantage. Forrester found that by supporting data science efforts with larger budgets and platforms (that serve to unify technology and infrastructure for knowledge management and collaboration), insight-driven businesses were *twice as likely* to achieve market-leading positions in their industries than technology laggards. Further, while insight-driven businesses constituted only twenty-two percent of the total firms being studied, Forrester predicted that they would earn $1.2 trillion in revenue in 2020, a prediction that has largely come to pass. (Microstrategy, 2020)

Perhaps the most significant way in which insights-driven firms differentiate themselves is through their ability to take advantage of the revolutionary technological advances of Industry 4.0. Since 2011, the term "Industry 4.0" has been used to describe smart manufacturing ecosystems that enhance intelligence, automation, and interconnectedness to solve the unique challenges of the contemporary marketplace. These are the same technologies used to support digital transformation outside of manufacturing. (Radziwill, 2020) This fourth industrial revolution has introduced intelligent cyber-physical systems that link people, objects, data sources, and systems using local and global networks. At the heart of Industry 4.0 is data.

As a result, as systems evolve in Industry 4.0 and digitally transformed organizations, testing needs to be examined from a broader perspective. Consequently, this article examines the test pyramid in the context of requirements for data and insights-driven businesses, in particular, to support testing for machine learning (ML) systems. We propose an adaptation of the test pyramid to help software and analytics professionals design quality into systems more holistically, thinking about software not only from the perspective of the applications and services, but also the data and models that supply them with information.

## The Test Pyramid

Although it is a familiar concept in software engineering, the test pyramid did not originate there. Perhaps the earliest explicit description of a testing pyramid was by Rouchon (1990). Rouchon's primary concern was establishing a way to measure quality in the construction of composite aircraft structures. Up to that point, aircraft manufacturers had relied on a single source for their materials. However, introducing composites presented several new quality challenges. First, the quality of composite parts is heavily process dependent, and the components develop their mechanical properties only after being processed. Second, the composite materials often come from a variety of sources instead of a single selected source. Finally, the diversity of different materials (e.g. metallic and non-metallic), and their interactions with each other, add complexity. (Goldbach, 2010) Rouchon's testing pyramid begins with the most fundamental components, the composite materials, at the bottom, and moves through the increasing complexity of the parts in combination until it culminates in entire systems at the top of the pyramid.

This pattern will be familiar to all software professionals, so it is not a surprise that the test pyramid would find applicability in software engineering. It came to wider prominence in Mike Cohn's 2009 book *Succeeding with Agile*, where it was initially referred to as the "Test Automation Pyramid". (Cohn, 2009) It is shown in Figure 1. One reason that the term "automated" might have fallen out of Cohn's original term "test automation pyramid" is that while automation is tacitly understood to be applicable at all levels of the pyramid, not every test *should* be automated. Some elements of UI, such as layout, are usually out of scope for the sometimes limited aesthetic judgment of automated tests and are best left to humans. (Vocke, 2018) As another example, exploratory testing (by definition) can not be prescribed.

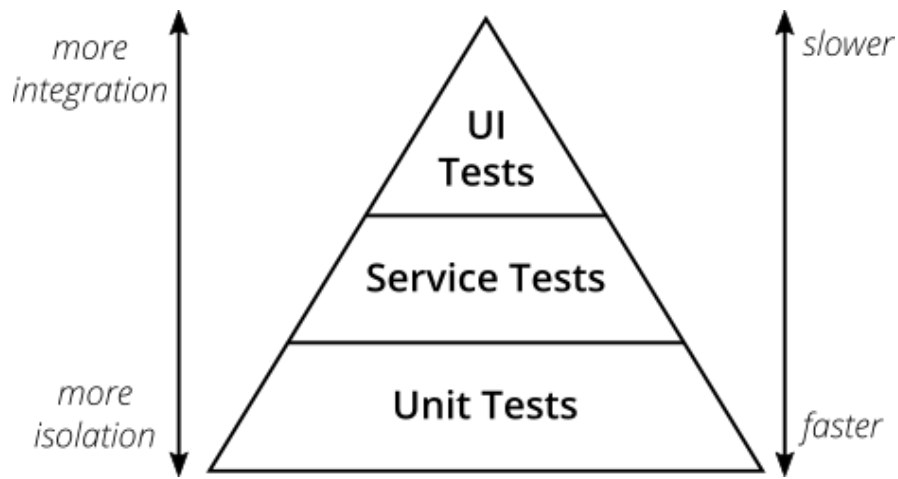

**Figure 1.** Cohn's (2009) test pyramid as described by Vocke (2018).

The bottom level of the pyramid consists of automated unit tests, which make up about 70% of the testing effort. (Oliinyk and Oleksiuk, 2019) Unit tests only require small amounts of code, are written by developers, and support frequent tests. They return quick results and are easier to create, manage, and maintain. The middle level consists of automated service tests for APIs and integration, and under ideal circumstances, should occupy about 20% of the testing effort. At this level, testing involves units in combination. These tests are therefore more complex to build and maintain, which means there should be fewer of them than there are unit tests. Finally, the top of the pyramid consists of automated UI tests, which occupy approximately 10% of the testing effort. These tests cover end-to-end workflows throughout the entire system. They can be slow and difficult to maintain, but often provide excellent insight into overall system function.

One possible variation of the test pyramid breaks the service level into three subsections for component tests, integration tests, and API tests. (Willett, 2016) Interactions between applications and databases are usually covered in the integration layer. Sometimes, the test pyramid is annotated with a "cloud" on top to reflect testing that can not be fully or partially automated (for example, exploratory tests and accessibility tests).

In practice, the pyramid tends to be inverted, with the more brittle and expensive UI tests appearing much more frequently. (Contan et al., 2018) In this "ice cream cone" model (Figure 2), there are far fewer automated unit tests indicated at the bottom, many more automated UI tests at the top, and a huge serving of manual tests at the top, with more of the testing responsibility pushed to the testers or QA team due to inadequate unit test coverage from developers. The ice cream cone is often the result of upper management being interested only in end-to-end testing that demonstrates the functionality of the product, and providing fewer resources for unit testing. Because of its nature, unit testing does not provide the evidence stakeholders sometimes need for sign-off. (Willett, 2016)

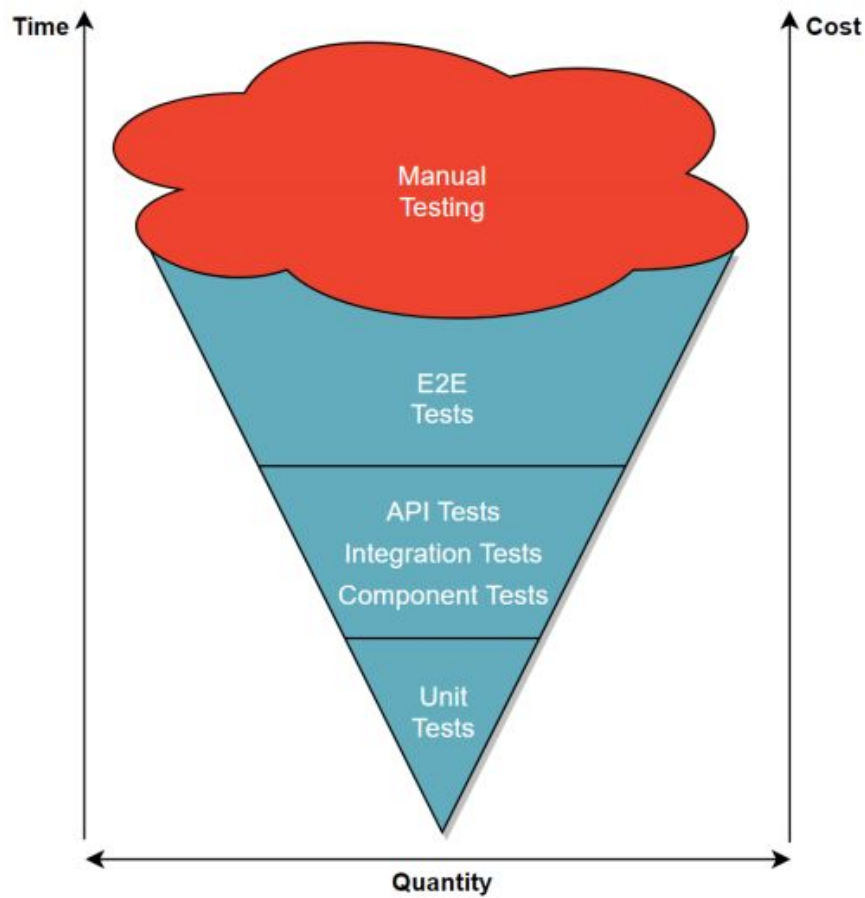

**Figure 2.** Testing in practice; the "inverted test pyramid" as described by Hartikainen (2020).

General practice is that low-level unit tests are easier to run, provide feedback more quickly, and can catch many issues early on. Ideally, high-level tests should not be relied upon to catch tests that should have been managed at the lower levels. (Oliinyk, 2019) Putting the weight for identifying problems in the high level tests can present several problems. In particular, if a UI does not allow a user to encounter certain situations, a high-level, end-to-end test will not identify it. Further, identifying the problem during the end-to-end could require an extensive examination of the entire system, a consequence that could have been eliminated with unit checks in the earlier stages. (Farias, 2019)

Some have criticized the test pyramid for not incorporating market risk into its scope to provide checkpoints that validate whether or not the project continues to be one that meets the strategic requirements of the organization. Others have pointed out that unit tests are not always useful for testing microservices, which are already small and require more testing relating to their interactions; in these cases, the pyramid may be more diamond-shaped as integration tests dominate. (Contan et al., 2018) There may also be multiple test pyramids: one for each microservice and one for the system in which they are embedded. (Wolff, 2016)

The progression of the levels on the test pyramid (from the bottom to the top) reflects the shift from testing *within the software components,* to tests *between software components*, and finally *across all components as a unified system.* Cohn suggests that a strong foundation requires extensive unit test coverage, and organizations should prioritize their testing investments accordingly; many others have echoed this. (Vester, 2017; Tucker et al., 2018; Nader-Rezvani & McDermott, 2019)

## Testing in Data Management and Data Science

While software development teams often reflect a mature application of the test pyramid and its variations, testing related to data and data management is usually performed independently. Data profiling, which is performed to assess quality, is an iterative process and often done as part of data wrangling. This "preparation, linkage, and exploration can easily consume the majority of the project's time and resources." (Keller et al., 2020) Testing for data science models and pipelines is often performed ad hoc, if at all, in part because models are often created by subject matter experts with experience in machine learning.

The criticality of the relationship between data testing and software testing was recognized as early as Bobrowski & Yankelevich (1998), who explained it this way:

> "Usually, testers and engineers assume that the data (in a production environment) is correct, and test the system considering its behavior. However, as we have said, this is not the case in the real world. When a new system is incorporated to an existing environment, the data it uses must be analyzed to understand its usefulness. Moreover, an old system may be using corrupted data. We believe that the verification of a system must include the verification of the data it works on. Besides, we believe that many testing techniques can be adapted in order to be used to test data."

These authors recommend articulating data quality requirements *for software*, and testing against those non-functional requirements with the same regard that might be afforded to other non-functional characteristics. In other words, software testing should require prequalification of data, if it is not to be done during the software testing process.

Data quality is often compromised by the application of too many technology tools, conflicting needs between diverse stakeholders, changing business requirements, and poor quality standards. Insights-driven firms need to close the data-sized gap in the test pyramid by building closed-loop systems to provide quality assurance for data science through a holistic approach. (Ereth, 2018) These gaps include ETL (Extract, Transform, and Load) testing, (Big) Data testing, machine learning (ML) testing, and data pipeline tests.

**ETL (Extract, Transform, and Load) Testing**

Testing ETL procedures occurs whenever source data is pulled from a repository, manipulated (or transformed) according to rules, and prepared for ingestion by other systems (like data marts supporting specific business lines). The approach is associated with quality assurance in a data warehouse. (El-Gamal et al., 2013)  Vyas & Vaishnav (2017) provided a comparative analysis of ETL testing types and the time commitment required for each. They include methods for validating the data in place as well as in motion (between source and target, or at the point of transformation):

- Production validation
- Source-to-target counting
- Source-to-target validation
- Application migration testing
- Checks on database constraints
- Duplicate checking
- Tests against specific data quality dimensions
- Regression testing for transformations

They highlight some challenges associated with ETL testing, for example that databases often contain incorrect, insufficient or duplicate data; that the volume of data can be too great for tests to be effective; and that ETL testers rarely have access to the source data (or the business context for the output) which can make robust testing difficult or impossible. Dakrory et al. (2015) point out that complexity in the data warehouse and cost of automation often leads to an over-reliance on manual testing. They recommend a battery of 14 tests (e.g. checking data types, appropriate formats, values within expected bounds) organized in terms of the 6 key data quality dimensions (accuracy, completeness, consistency, timeliness, uniqueness, and validity).

**(Big) Data Testing**

Not all data is stored in data warehouses where it conforms to a predetermined schema. Many datasets are stored on cloud service platforms that abstract the details of infrastructure management and allow users to more easily "move the compute to the data." (Wright et al., 2015) This makes it possible for software and models deployed in the cloud to more easily scale to ever-increasing amounts of data. The data itself is also no longer constrained to a schema, as it was in the traditional data warehouse, but can be stored in different ways:

- **Structured** - relational databases (RDBMS), enterprise software databases (e.g. CRM, ERP)
- **Unstructured** - images, audio, video, HTML pages, logs, telemetry
- **Semi-structured** - JSON, XML, CSV formats

Punn et al. (2019) describe Big Data testing as a superset of ETL testing that also includes systematizing tests for unstructured and semi-structured data, and potentially transforming unstructured data to semi-structured data in order to conduct tests. In addition, tests need to be adapted to accommodate the schema-on-load approach, in which the structure of the data is not defined before it is stored. Combining these concerns, these authors outline three stages of Big Data testing:

- **Data staging validation** - validating the data sourced from different origins
- **Procedure validation** - analyzing the legitimacy of business logic at each step
- **Output validation** - verifying that processed files have target characteristics

Sneed & Erdoes (2015), describing a case study that integrated these concepts, highlighted the importance of test automation to validate Big Data, noting that the "sheer volume of data together with the variety of the data makes it impossible to test manually." More significantly, they identified that the most time consuming part of their Big Data testing process was validating and reconstructing the mapping rules, a process that requires domain knowledge. They concluded that *testing Big Data must incorporate greater engagement from domain experts*, and that it will evolve to be more qualitative than quantitative.

## Machine Learning (ML) Testing

Machine learning (ML) systems have emergent properties. Quality assurance and testing can only be performed by considering the system as a whole because it is no longer practical to articulate deterministic test cases. In addition, in many cases it is not as apparent how to break an ML system into smaller components to test as units, meaning that testing challenges emerge at integration and system levels. Zhang et al. (2020) present models to help software engineers, domain experts, and quality professionals begin planning and executing tests on ML systems more effectively. This requires a strategically selected combination of:

- **Offline** tests (prior to model deployment) and **online** tests (once model is deployed)
- Bug detection in **data**, the **learning process**, and ML **frameworks**

Although there are some tools available to support unit tests for models that use ML frameworks, the status quo for ML testing consists of heterogeneous tools and fractured approaches, many of which are still at the research stages. Zhang et al. (2020) also note that the testing responsibility may also begin to spread across many roles, which may necessitate a new approach to testing to more fully involve domain experts in the process.

### Data Pipeline Testing

Data pipelines automate one or more steps (acquire data, wrangle data, explore data, create model, validate model, deploy model) in a data science workflow. Although the concept of automated data pipelines is not new (e.g. Crossley et al., 2008) the state of professional practice is just beginning to form. (e.g. Atwal, 2020) One technique is to test data pipelines on simulated data that has the same properties and characteristics the real data will have. (Weilbacher et al., 2020) Automated ML (or "AutoAI") systems, that can autonomously ingest and pre-process data, and create new features and scoring systems based on target requirements, are also considered to be special cases of data pipelines. Although many data scientists feel that these systems will be inevitable in the workplace, feelings are mixed regarding their utility and trustworthiness. (Wang et al., 2019)

Although the gap between quality assurance practice for data versus software has long been acknowledged (e.g. Kumar, 2019) there are certainly methods in place for testing in data management. Similarly, data scientists may be incorporating unit tests into their model code, and data engineers may be running integration tests to confirm the performance of their pipelines. What is missing is an organizational view of these testing efforts that acknowledges each of them are critical for producing and delivering information in an insights-driven business.

## A Holistic View of Testing for Industry 4.0

Testing is, by definition, risk based. It acknowledges the possible impact of uncertainty and strives to anticipate and mitigate it before it becomes a threat or yields a negative outcome. However, risk-based testing provides a more explicit accounting for risk at every step of the testing process. This is particularly important for products and processes that have a high cost of failure. (Bach, 2012; Felderer & Schieferdecker, 2014; Großmann et al, 2019)

Incorporating risk-based testing to manage quality throughout the entire information supply chain, and addressing each of the areas outlined above, begins to approximate Juran's Quality by Design principle and extends it into a new domain. (Juran, 1992) Foidl (2019) took one step towards this end by constructing a framework for data validation based on the risk of poor data quality, specifically tailored to the context of ML testing. Shown in Figure 3, this captures the two categories that require tests (data pipelines and ML models) and shows how the components relate to deployment infrastructure, ML frameworks, and traditional software:

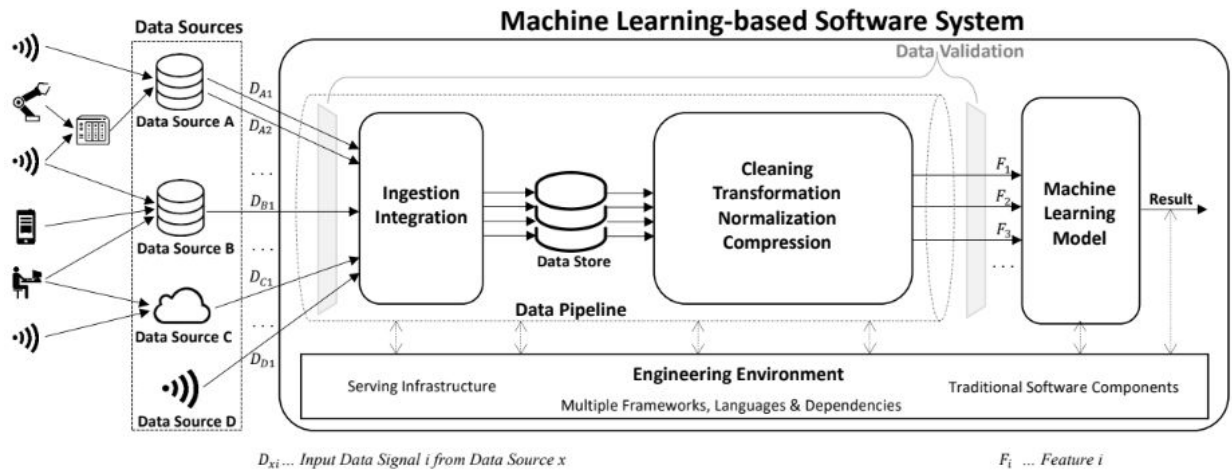

**Figure 3.** Mechanics of online and offline ML testing, from Foidl (2019)

The traditional domain of ETL testing for data warehouses (supplemented by different modes of Big Data testing for unstructured and semi-structured data), coupled with ML systems testing at the level of the models and pipelines, thus overlaps with traditional software testing at the interface of the ML model and the systems that work with them or consume and display their insights. As a result, we propose an evolution of the test pyramid that captures the increasing dependence of software systems on the underlying data and models, and does not treat ML testing or Big Data testing as unique, separate activities.

The proposed pyramid (Figure 4) centers the test strategy around where the tests are performed, not what kinds of tests are created. It assumes that unit testing will be performed at all levels of the pyramid, supplementing functional tests. It does not distinguish between a user who is a human and a user who is a system, acknowledging that uncertainty is likely to arise in both cases given the increasing complexity of data-driven systems. It also removes the potential perception that because unit tests should provide the foundation for system behavior, performance, and stability, that they can stand alone and do not need to be supplemented by contract or functional tests at all levels; in fact, the need for functional testing is only increasing. (Zeller & Weyrich, 2016; Mirza & Khan, 2018; Liu & Nakajima, 2020)

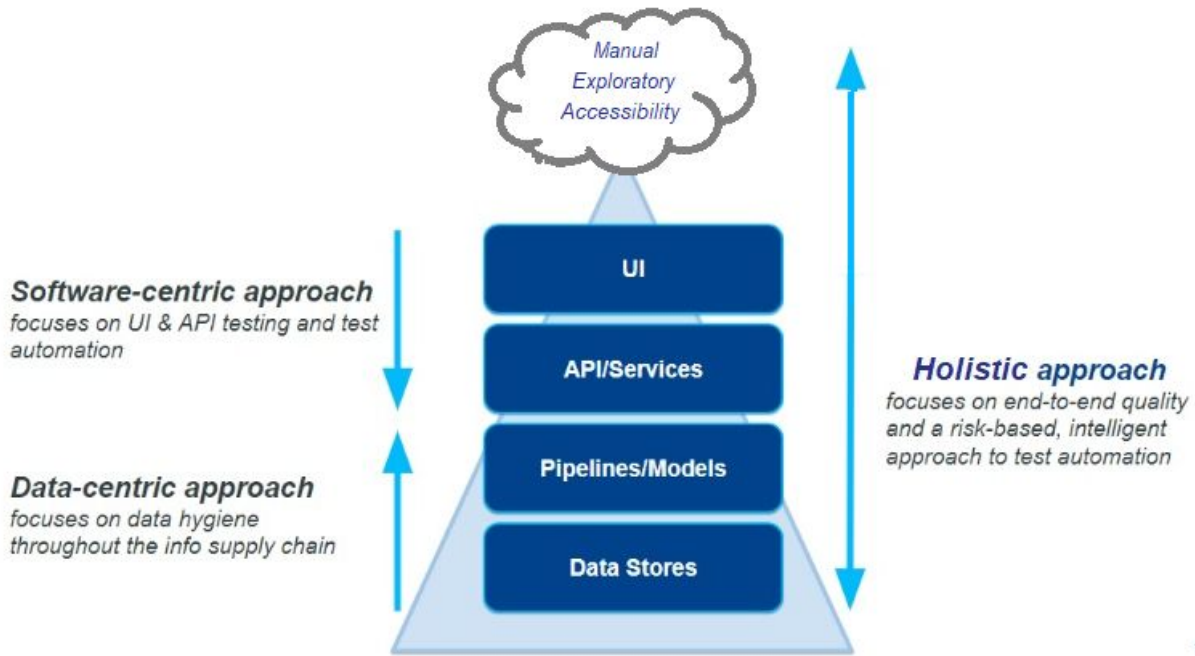

**Figure 4.** A holistic test pyramid that bridges data and traditional software testing.

Although this model has not been empirically validated, its utility has been indicated in limited settings. For example, in April 2020, an insurance company wanted to prioritize testing activities because their budget had been recently punctuated by the global COVID-19 epidemic. They conducted a high-level failure mode effects analysis (FMEA) on six key business processes. Both the Chief Information Officer (CIO) and Chief Data/Analytics Officer (CDAO) had representatives at this workshop with decision-making authority. They collected data about the frequency of internal failures (experienced by their employees, and captured in bug reports) and external failures (that impacted their customers, captured in bug reports and customer complaints). By reflecting on the severity, occurrence, and ability to detect different failure modes, and how the risk priority would change if they added tests at any of the four levels in Figure 3, they recognized that enhanced testing at the data and pipeline levels would substantially decrease the risk of automated loan processing while reducing customer-facing external failures.

## Conclusions

For Industry 4.0 and in all digitally transformed organizations, the quality of software can be strongly affected by the quality of the data, pipelines, and models supporting it. Although data quality and software quality have traditionally been managed by different people in different functional areas of the organization, taking a more holistic view of testing may help drive efficiencies in the future. We encourage organizations to consider planning their investments in testing using a framework that treats data, pipelines and models, services, and interfaces from a holistic perspective that puts business processes and risk at the center. Furthermore, we encourage research that reflects the increasing interconnectedness of systems through this relationship.

## References

Atwal, H., 2020. Build Feedback and Measurement. In Practical DataOps (pp. 113-137). Apress, Berkeley, CA.

Bach, J., 1999. Heuristic risk-based testing. Software Testing and Quality Engineering Magazine, 11(9).

Bobrowski, M., Marré, M. and Yankelevich, D., 1998. A software engineering view of data quality. Proceedings of Second International Software Quality in Europe.

Cohn, M., 2009. Succeeding with agile: software development using Scrum. Pearson Education. p. 312-314.

Contan, A., Dehelean, C. and Miclea, L., 2018, May. Test automation pyramid from theory to practice. In 2018 IEEE International Conference on Automation, Quality and Testing, Robotics (AQTR) (pp. 1-5). IEEE.

Crossley, J.H., Sjouwerman, L.O., Fomalont, E.B. and Radziwill, N.M., 2008, July. NRAO VLA archive survey. In Observatory Operations: Strategies, Processes, and Systems II (Vol. 7016, p. 70160O). International Society for Optics and Photonics.

Dakrory, S.B., Mahmoud, T.M. and Ali, A.A., 2015. Automated ETL testing on the data quality of a data warehouse. International Journal of Computer Applications, 131(16), pp.9-16.

El-Gamal, N., ElBastawissy, A. and Galal-Edeen, G., 2013, March. Data warehouse testing. In Proceedings of the Joint EDBT/ICDT 2013 Workshops (pp. 1-8).

Farias, J. 2019. Comment on *Software Quality and Assurance Testing.* https://sqa.stackexchange.com/questions/37623/is-inverted-test-pyramid-really-anti-pattern. Retrieved July 3, 2020.

Fenty, K., 2016. Data science platforms help companies turn data into business value. *A Forrester Consulting Thought Leadership Paper Commissioned by DataScience.* [1-114Y3BP]

Felderer, M. and Schieferdecker, I., 2014. A taxonomy of risk-based testing.

Goldbach, S., Franke, R., & Dresden, I. M. A. Materials and NDT Applications in Aircraft Component Development Tests.
http://www.ndt-aerospace.com/Portals/aerospace2010/BB/we4b2.pdf

Großmann, J., Felderer, M., Viehmann, J., & Schieferdecker, I. (2019). A Taxonomy to Assess and Tailor Risk-Based Testing in Recent Testing Standards. *IEEE Software*, *37*(1), 40-49.

Hartikainen, V., 2020. Defining suitable testing levels, methods and practices for an agile web application project. Dissertation, Lappeenranta-Lahti University of Technology (LUT) School of Engineering.

Juran, J. M. (1992). *Juran on quality by design: the new steps for planning quality into goods and services*. Simon and Schuster.

Keller, S.A., Shipp, S.S., Schroeder, A.D. and Korkmaz, G., 2020. Doing Data Science: A Framework and Case Study. Harvard Data Science Review, 2(1).

Kumar, S., 2019. Testing Improvement in Business Intelligence Area. *Research proposal.*

Liu, S. and Nakajima, S., 2020. Automatic Test Case and Test Oracle Generation based on Functional Scenarios in Formal Specifications for Conformance Testing. IEEE Transactions on Software Engineering.

Microstrategy, 2020. Global State of Enterprise Analytics: Minding the Data-Driven Gap. Available from
https://www.microstrategy.com/getmedia/db67a6c7-0bc5-41fa-82a9-bb14ec6868d6/2020-Global-State-of-Enterprise-Analytics.pdf

Mirza, A.M. and Khan, M.N.A., 2018. An automated functional testing framework for context-aware applications. IEEE Access, 6, pp.46568-46583.

Morton, S.D., 1999. Bottoms Up! Testing Top-Down Software Designs (No. 1999-01-0953). SAE Technical Paper.

Nader-Rezvani, N. & McDermott, 2019. Executive's Guide to Software Quality in an Agile Organization. Springer.

Oliinyk, B. and Oleksiuk, V., 2019, November. Automation in software testing, can we automate anything we want. In Proceedings of the 2nd Student Workshop on Computer Science & Software Engineering (CS&SE@ SW 2019), Kryvyi Rih, Ukraine (pp. 224-234).

Punn, N.S., Agarwal, S., Syafrullah, M. and Adiyarta, K., 2019, September. Testing Big Data Application. In 2019 6th International Conference on Electrical Engineering, Computer Science and Informatics (EECSI) (pp. 159-162). IEEE.

Radziwill, N., 2020. Connected, Intelligent, Automated: The Definitive Guide to Digital Transformation and Quality 4.0

Rouchon, J., 1990. Certification of large airplane composite structures. In ICAS Congress Proceedings (Vol. 2, pp. 1439-1447).

Sneed, H.M. and Erdoes, K., 2015, April. Testing big data (Assuring the quality of large databases). In 2015 IEEE Eighth International Conference on Software Testing, Verification and Validation Workshops (ICSTW) (pp. 1-6). IEEE.

Tucker, H., Hochstein, L., Jones, N., Basiri, A. and Rosenthal, C., 2018. The business case for chaos engineering. IEEE Cloud Computing, 5(3), pp.45-54.

Vester, J., 2017. RESTful API Lifecycle Management. URL: https://dzone.com/storage/assets/4960646-dzone-rc238-restfulapilifecyclemanagement.pdf (accessed on Mar. 21, 2018).

Vocke, H, Feb. 26, 2018. The Practical Test Pyramid. https://martinfowler.com/articles/practical-test-pyramid.html. Retrieved July 3, 2020.

Vyas, S. and Vaishnav, P., 2017. A comparative study of various ETL process and their testing techniques in data warehouse. Journal of Statistics and Management Systems, 20(4), pp.753-763.

Wang, D., Weisz, J.D., Muller, M., Ram, P., Geyer, W., Dugan, C., Tausczik, Y., Samulowitz, H. and Gray, A., 2019. Human-AI Collaboration in Data Science: Exploring Data Scientists' Perceptions of Automated AI. Proceedings of the ACM on Human-Computer Interaction, 3(CSCW), pp.1-24.

Weilbacher, P.M., Palsa, R., Streicher, O., Bacon, R., Urrutia, T., Wisotzki, L., Conseil, S., Husemann, B., Jarno, A., Kelz, A. and Pécontal-Rousset, A., 2020. The Data Processing Pipeline for the MUSE Instrument. arXiv preprint arXiv:2006.08638.

Willett, J. 2016. The Evolution of the Testing Pyramid. https://www.james-willett.com/the-evolution-of-the-testing-pyramid/

Wolff, E., 2016. Microservices: flexible software architecture. Addison-Wesley Professional.

Wright, N.J., Dosanjh, S.S., Andrews, A.K., Antypas, K.B., Draney, B., Canon, R.S., Cholia, S., Daley, C.S., Fagnan, K.M., Gerber, R.A. and Gerhardt, L., 2015. Cori: A pre-exascale supercomputer for big data and hpc applications. Big Data and High Performance Computing, 26, p.82.

Zeller, A. and Weyrich, M., 2016, September. Challenges for functional testing of reconfigurable production systems. In 2016 IEEE 21st International Conference on Emerging Technologies and Factory Automation (ETFA) (pp. 1-4). IEEE.

Zhang, J.M., Harman, M., Ma, L. and Liu, Y., 2020. Machine learning testing: Survey, landscapes and horizons. IEEE Transactions on Software Engineering.